# Crossover from meta-magnetic state to spin-glass behaviour upon Ti-substitution for Mn in $CuMn_2O_4$

P. Patra[1], I. Naik[1*], S. D. Kaushik[2] and S. Mohanta[1]

[1]Department of Physics, North Orissa University, Baripada – 757003 (INDIA)

[2]UGC-DAE Consortium for Scientific Research Mumbai Centre, BARC, Trombay, Mumbai – 400085 (INDIA)

*Corresponding author: indrajit_naik@yahoo.co.in

**Abstract:** Tetragonal distorted spinel of $CuMn_{2-x}Ti_xO_4$ (x = 0, 0.25 and 0.50) was prepared by solid state reaction method followed by neutron diffraction, FTIR spectroscopy, dielectric spectroscopy and magnetization measurements. The neutron diffraction and FTIR spectroscopy provide the information regarding phase formation. But, the magnetic susceptibility clearly shows the ferri-magnetic order below 76K associated with metamagnetic state which turns to spin-glass behaviour upon Ti-substitution for Mn in $CuMn_2O_4$. In addition, the room temperature M(H) of all the spinels are described by the Arrot's plot for weak ferromagnetism. Further, the room temperature dielectric measurement provides the electronic property of $CuMn_2O_4$ dominated by grain boundary effect that is significantly increased upon Ti-substitution for Mn in $CuMn_2O_4$.

## 1. Introduction

The oxide spinel's are often represented by the general formula $AB_2O_4$ which crystallizes as FCC lattice with cubic closed packed of oxide anions and space group *Fd3m*. Further, spinels can be divided into three types according to their cationic distributions among the tetrahedral and octahedral sites as $\{A_{1-x}B_x(B_{2-x}A_x)O_4\}$ which provides (a) normal spinel for $\{x=0\}$, (b) inverse spinel for $\{x=1\}$ and (c) mixed spinel for $\{0<x<1\}$. Nowadays, spinel materials are becoming very important for condensed matter physics research group's across the world due to their multifunctional properties and technological applications [1–4]. One such spinel compound is $CuMn_2O_4$ in which more than one oxidation states occur in Manganese (Mn) and Copper (Cu) and their distributions among tetrahedral and octahedral sites depend strongly on the preparation conditions, thus it's physical properties vary [5–9]. As a consequence, it creates a center of attention for the scientific community worldwide in condensed matter physics. But, most of the results in $CuMn_2O_4$ are reported on tetragonal symmetry which is obtained due to the coherence in degenerate energy levels of Cu and Mn result the change in cubic symmetry of $CuMn_2O_4$ to tetragonal symmetry [5, 10, 11]. At the other side, if the distortion at B-site corresponds to $c/a>1$ compensate by opposite distortion at A-site corresponds to $c/a<1$ then cubic symmetry of $CuMn_2O_4$ can be retained [12]. Besides the effect of site distribution on multivalent cations of Cu and Mn, a significant change in magnetic and dielectric properties has been reported by substituting foreign cations irrespective of the crystal symmetry. As per the available literature on the title compound, most of the scientific groups have made an attempt to study $CuMn_2O_4$ by substituting foreign atoms (Ni and Zn) in tetrahedral site with copper resulting considerable change in structure and electrical behavior whereas Cd-substitution shows BB-interaction is responsible for the reduction of ferromagnetic behavior [12-15]. But, the octahedral site substitution by foreign atoms (Al and Cr) with manganese provides clusterization of octahedral J-Teller ions leading to tetragonal distortion whereas Fe-substitution provides increase of catalytic activity [16, 17].

In this paper, we have reported the room temperature neutron diffraction, FTIR spectroscopy, complex dielectric spectroscopy and temperature dependent magnetization on $CuMn_{2-x}Ti_xO_4$. The, Titanium (Ti) substitution at Mn-site of $CuMn_2O_4$ is done, here, in-order to observe the change in metamagnetic and dielectric behavior as a consequence. But, the neutron diffraction data analysis assists us in understanding the FTIR spectroscopy and electrical properties in terms of grain and grain boundary through dielectric study. Further, the magnetization is analyzed to explain adequately the magnetic behavior of $CuMn_{2-x}Ti_xO_4$. From the above physical property measurements, we observed that the metamagnetic behavior of $CuMn_2O_4$ is turned out to spin-glass behavior upon substitution of Ti for Mn-site. If the atomic moments are aligned anti-parallel with weak antiferromagnetic interactions, then metamagnetic state is formed which can be changed to weak ferromagnetic or ferromagnetic or anti-ferromagnetic when high external magnetic field is applied. On the contrary, spin-glass state is a disordered magnetic state in which spins are not aligned in a regular pattern but contains mixture of ferromagnetic and anti-ferromagnetic bonds which are arranged randomly that reveals frustrated interaction.

## 2. Experimental details

The $CuMn_{2-x}Ti_xO_4$ with x = 0, 0.25, 0.5 were prepared by mixing CuO, TiO and $MnO_2$ at appropriate molar proportion and calcinated at $900^0$C for 50hrs. Subsequently, the calcinated powder was grinded through an agate mortar pestle for homogenization by adding small amount of acetone as volatile substance. This is followed by making pellets through a hydraulic press and sintered at $950^0$C for 72hrs by an aluminium crucible for densification.

To identify the phase purity of the above materials, the neutron diffraction and FTIR-spectra measurements were carried out at room temperature. The Neutron diffraction (ND) measurement was done within 10 – 120 degree scattering angle with neutron wavelength of 1.48 Å using multi position-sensitive detector (PSD) based focusing crystal diffractometer at the National Facility for Neutron Beam Research (NFNBR), Dhruva reactor, Mumbai (India) set up by UGC-DAE CSR Mumbai Centre [18]. Subsequently FTIR-spectra measurements were done within the spectral range of 400-4000 cm$^{-1}$ using Germanium coated KBr beam splitter and DLATGS detector at a resolution of 4 cm$^{-1}$. In addition, the magnetic and complex dielectric spectroscopy measurements were also presented on these materials. Here, the magnetic measurement was carried out in the temperature range 2K to 300K using M/s. Quantum Design USA make cryogen free Helium Re-liquefier based 9T Physical Property Measurement System (PPMS) in vibrating sample magnetometer (VSM) mode. But the dielectric spectroscopy measurement was carried out at 300K over the frequency range 60 Hz to 2 MHz for an oscillation voltage of 1.0 V with the help of precision LCR meter (model E4980A, M/s. Keysight, USA).

## 3 Results and discussions

### 3.1 Crystal structure analysis

*Neutron Diffraction*: The room temperature neutron diffraction (ND) patterns of $CuMn_{2-x}Ti_xO_4$ (X=0, 0.25 and 0.50) followed by their analysis using Rietveld method and employing Fullprof suite programme are explained here in detail. Indeed, all the materials were well fitted by distorted tetragonal symmetry of space group *I41/amd* spinel phase with cell volume 286.522Å$^3$ (with cell parameter a =5.871 Å, c = 8.313 Å) along with traces of secondary phase $Mn_3O_4$ with same space group symmetry as that of parent material i.e. *I*41/*amd*, but with relatively larger cell volume 310.547 Å$^3$ (with cell parameter a = 5.741 Å and c = 9.428 Å), this secondary phase has been a part of the parent system in most of the previous reports [9, 19, 20]. The room temperature neutron diffraction data have been reported elsewhere in brief [21]. Here the detailed and augmented neutron diffraction data analysis is presented in fig 1, which includes the disorder particulars, cationic distribution, and the bond distance and angle on $CuMn_{2-x}Ti_xO_4$ series of material. The goodness of the fitting parameters is also mentioned in the figure 1.

In present work the impurity phase of $Mn_3O_4$ is around 15% in $CuMn_2O_4$, and it remains more or less constant with Ti-substitution at x = 0.25. But with further Ti-substitution, the un-reacted $TiO_2$ is observed at x = 0.5, though $Mn_3O_4$ phase seems to have decreased from around 15% to 8% and with additional phase of $TiO_2$ (~ 4-6%). In addition, the primary phase exhibits significant change in crystal cell parameters that is *a-axis* increases and *c-axis* decreases upon Ti-substituted derivatives as compared to $CuMn_2O_4$ (see table-I). Here we also notice that overall cell volume increases, pre-dominantly because of large $Ti^{4+}$ ion substitution. It is worth mentioning here that the goodness of fitting parameter improved upon considering the distortion in the material, i.e. Cu and Mn atoms were found on crossover site. This implies that disorder has been intensified upon the Ti-substitution, as Ti being in the 4+ valance state which likely to create $Mn^{3+}$ and $Mn^{4+}$ and correspondingly $Cu^{1+}$ and $Cu^{2+}$ thus facilitating to intensify the disorder in the material. The details of disorder for the series of material are described in table-II. Further, the presence of $TiO_2$ as unreacted phase indicates the breaching of solubility limit in $CuMn_{1.50}Ti_{0.50}O_4$. Here we must mention that as a consequence of $Ti^{4+}$ substitution, the microscopic variation in the materials are observed which are reflected in terms of the bond lengths and bond angle variation. These variation due to Ti-substitution at Mn-site are determined and it is observed that the obtained bond lengths for octahedral (tetrahedral) sites are Mn–O = Cu–O = 1.990Å and 1.923Å (1.999Å) in $CuMn_2O_4$, Mn–O = Ti–O = Cu–O = 1.998Å and 1.917Å (2.002Å) in $CuMn_{1.75}Ti_{0.25}O_4$ and Mn–O = Ti–O = Cu–O = 2.010Å and 1.916Å (2.010Å) in $CuMn_{1.50}Ti_{0.50}O_4$. From the zoomed region of fig. 1, plotted separately in the right side, we can also see the variation in the 100% peak, the two distinct peaks for $CuMn_2O_4$ merges in to one upon Ti-substitution, which also signifies the reduction of $Mn_3O_4$ concentration (~15% to ~8%) as the left-hand side peak in the doublet is predominantly has a contribution from $Mn_3O_4$. Conversely, the formation of phase on $CuMn_{2-x}Ti_xO_4$ was also confirmed by room temperature FTIR spectroscopy analysis and the calculated bond length of FTIR spectroscopy are compared with the obtained bond length of Rietveld analysis.

*FTIR Spectroscopy*: In order to verify the phase purity of $CuMn_{2-x}Ti_xO_4$, the room temperature FTIR-spectra were obtained within the spectral range of 400-4000 $cm^{-1}$ as shown in fig. 2. A total of 20 scans were co-added at a resolution of 4 $cm^{-1}$ using Germanium coated KBr beam splitter and DLATGS detector. However, the position of all the absorption bands except 633 $cm^{-1}$ and 973 $cm^{-1}$ looks like similar implies the formation of same phase associated with the modification of bond lengths due to Ti-substitution in $CuMn_2O_4$. Because, upon Ti-substitution, the intensity of 973 $cm^{-1}$ decreases and 633 $cm^{-1}$ shifts towards the higher wave number side significantly. Within the spectral range of 500-1000 $cm^{-1}$ which is commonly known as finger print region, $CuMn_2O_4$ shows four absorption bands at 510 $cm^{-1}$, 605 $cm^{-1}$, 633 $cm^{-1}$ and 973 $cm^{-1}$. However, the bond lengths of Mn–O = 2.078Å and 2.017Å and Cu–O = 2.132Å in $CuMn_2O_4$, Mn–O (Ti–O) = 2.090Å (2.113Å) and 1.895Å (1.956Å) and Cu–O = 2.124Å in $CuMn_{1.75}Ti_{0.25}O_4$ and Mn–O (Ti–O) = 2.097Å (2.120Å) and 1.739Å (1.759Å) and Cu–O = 2.113Å in $CuMn_{1.50}Ti_{0.50}O_4$ are calculated from three absorption bands that lies between 500 – 800 $cm^{-1}$ using $k = 24/r^3$ for Mn-O/Ti-O and $k = 19/r^3$ for Cu-O with $k = (2\pi\bar{\nu}c)^2 \mu$, where $\mu$ and $r$ are the reduced mass and average bond length between two atoms respectively. These bond lengths are quite consistent with the obtained bond lengths from Rietveld analysis of neutron diffraction irrespective of the coordination in $CuMn_{2-x}Ti_xO_4$.

### 3.2 Magnetic Measurements

The ZFC and FC magnetization measurements on $CuMn_{2-x}Ti_xO_4$ were carried out at 500 Oe magnetic fields within the temperature range 2 – 300K and depicted in fig. 3. Here, the FC magnetization of $CuMn_2O_4$ shows independent of temperature on cooling from 300K down to 80K below which it rises rapidly may be due to spin-canting and exhibits an anomaly at 24K and then increases very slowly on further cooling. However, the temperature dependent ZFC magnetization was found identical with FC magnetization above 56K below which ZFC bifurcates from FC behaviour and attend a broad maximum around 24K. Indeed, the temperature for the onset of bifurcation between ZFC and FC is ascribed to metamagnetic state which we found at 56K in our present study is slightly higher than the previous result may be due to the phase purity and different cationic distributions [9]. Further, our magnetic behaviour on $CuMn_2O_4$ is quite similar with the previous results of $CuMn_2O_4$ with tetragonal symmetry and cubic symmetry [5, 9, 22]. Upon Ti-substitution, the ZFC and FC magnetic behaviours are significantly altered particularly below 80K as can be seen in fig. 3. In both the derivatives, the onset temperature for rapid rise in magnetization may be due to spin-canting is lowered followed by shifting the temperature for bifurcation between ZFC and FC towards the 24K anomaly. Most interestingly, the rate of increase in FC magnetization slows down below 24K upon increasing Ti-concentration and becomes completely flattened in $CuMn_{1.50}Ti_{0.50}O_4$. This type of behaviour such as the bifurcation between ZFC and FC magnetization below which the FC magnetization becomes completely flattened and ZFC magnetization decreases non-exponentially are frequently obtained only in those materials where competition between anti-ferromagnetic and ferromagnetic ordering occurs resulting a spin-glass state [23]. The inset fig. 3 shows the temperature variation of inverse susceptibility as parabolic above 80K for all the materials implies the existence of ferrimagnetic behaviour due to the anti-parallel ordering of nearest neighbour magnetic moments with unequal magnitude. At room temperature, the inverse susceptibility of $CuMn_{1.75}Ti_{0.25}O_4$ decreases by a factor of approximately half as compared to $CuMn_{1.50}Ti_{0.50}O_4$ which shows very close to $CuMn_2O_4$. Here, the increase of inverse susceptibility in $CuMn_{1.50}Ti_{0.50}O_4$ is due to the compensation of paramagnetic behaviour with diamagnetic behaviour of impurity phase $TiO_2$ [24]. Further, the possible causes behind it is explained more elaborately using M(H) data in the last paragraph of this sub-section. In general, the ideal Curie-Weiss behaviour in the paramagnetic region denotes the non-interacting spins. Therefore, we have fitted the high temperature susceptibility (75-300K for $CuMn_2O_4$ and $CuMn_{1.50}Ti_{0.50}O_4$ and 75-175K for $CuMn_{1.50}Ti_{0.50}O_4$) by Curie-Weiss law $\{\chi(T)-\chi_0\}=C/(T-\theta)$ as can be seen in inset fig. 4 to obtain the effective magnetic moments ($\mu_{eff}$), Curie-Weiss temperature ($\theta$), van-vleck paramagnetism ($\chi_0$) and depicted in table–I. One can clearly notice from table-I that $\mu_{eff}$ decreases upon Ti-substitution in $CuMn_2O_4$. This is ascribed with distributions of magnetic ions which provides low $\mu_{eff}$ for $CuMn_{1.75}Ti_{0.25}O_4$ compared to other two as a consequence. Further, the frustrated parameters $f=|\theta|/T_N$ are calculated $\sim 1.33$, $1.37$ and $2.8$ for $CuMn_2O_4$, $CuMn_{1.50}Ti_{0.50}O_4$ and $CuMn_{1.50}Ti_{0.50}O_4$ respectively due to the random mixture of ferromagnetic and anti-ferromagnetic bonds in the materials.

Subsequently, we have derived a normalised susceptibility equation using the Curie constant ($C$) and Curie-Weiss temperature ($\theta$) as follows,

$$\frac{C}{\chi|\theta|} = sign(\theta) + \frac{T}{|\theta|} \quad (1)$$

This equation provides the magnetic interactions of the material which evolves with compositions. As can be seen from fig. 4, the $C/\chi|\theta|$ vs. $T/|\theta|$ provides the occurrence of short-range interactions in all the materials because, depending on the ratio of magnetic ordering temperature to Curie-Weiss temperature the deviation in Curie-Weiss behaviour corresponds to either short-range ($T/|\theta| \geq 1$) interaction or long-range ($T/|\theta| << 1$) interaction. In addition, all are showing downward deviation implies uncompensated interactions because upward deviation will give compensated interactions.

The variation of magnetization with applied field at 3K and 300K are depicted in fig. 5 and inset fig. 5 respectively. The complete hysteresis behaviours for all the materials have been reported elsewhere [21]. Therefore, here, we have given emphasis to only correlate $M(H)$ with $M(T)$ behaviour. At 3K, $CuMn_2O_4$ shows a hysteresis loop at I-quadrant quite similar with earlier reports of $CuMn_2O_4$ measured at 5K [5]. This type of behaviour is frequently obtained in $M(H)$ due to the co-existence of spin-canting and metamagnetic state and we claim the same mechanism here also. Upon Ti-substitution, the anomaly related to weak ferromagnetic in metamagnetic state washes out with broadening of coercivity along with the reduction of unsaturated magnetization compared to $CuMn_2O_4$. This is expected to be due to the strong interactions between ferromagnetic and anti-ferromagnetic moments which give rise to spin-glass state. But at 300K, $CuMn_{1.75}Ti_{0.25}O_4$ provides a kink around the origin of $M(H)$ like a small hysteresis loop with significant large magnetization value in comparison to $CuMn_2O_4$ and $CuMn_{1.50}Ti_{0.50}O_4$ associated with unsaturated magnetization above 0.5kOe field for all the materials. To explain the room temperature $M \sim H$ behaviour, we have plotted $M^2$ vs. $H/M$ which is commonly known as Arrott plot in the Stoner-Wohlfarth model of itinerant electron magnetism because, Arrott plot is an alternative method to investigate the weak ferromagnetic for a material. Inset fig. 6 clearly shows the variation of $M^2$ vs. $H/M$ with shifting of $M^2(H/M)$ towards left in $CuMn_{1.75}Ti_{0.25}O_4$ and towards right in $CuMn_{1.50}Ti_{0.50}O_4$ associated with curvatures in $M^2(H/M)$ curve as compared to $CuMn_2O_4$ in which a vertical straight line is obtained. Generally, the straight-line behaviour in $M^2$ vs. $H/M$ corresponds to homogeneous system without localized moments of spin and curvature corresponds to inhomogeneous system with localized moments. Further, if the tangent of the curve intercepts at positive $M^2$-axis implies weak ferromagnetic whereas magnetic phase transition occurs when the tangent intercepts at origin. As all the materials contain van-vleck paramagnetism, the $M^2(H/M)$ curves can't be understood easily in the present form. Therefore, Arrott plot is re-plotted as $(M-M_0)^2$ vs. $H/(M-M_0)$ after subtracting the magnetization due to van-vleck contribution ($M_0$) and shown in fig. 6. It reveals that all the materials belong to system of inhomogeneous and weak ferromagnetic due to the localized moments which increases upon Ti-substitution for Mn in $CuMn_2O_4$. However, the curve of $CuMn_{1.75}Ti_{0.25}O_4$ intercepts the $(M-M_0)^2$–axis at large positive value implies that the weak ferromagnetic order takes place over large scale as compared to $CuMn_2O_4$. Although the frustrated parameter is very large in $CuMn_{1.50}Ti_{0.50}O_4$ as compared to $CuMn_2O_4$ and $CuMn_{1.25}Ti_{0.75}O_4$, the curve of $CuMn_{1.50}Ti_{0.50}O_4$ intercepts $(M-M_0)^2$–axis at lower positive value very close with $CuMn_2O_4$ due to the additional impurity phase of $TiO_2$ which shows diamagnetic behaviour [24]. Due to the increased weak ferromagnetic in the materials, the anti-ferromagnetic moments interact strongly with ferromagnetic and thus anti-ferromagnetic order renormalizes the ferromagnetic order in such a way that the ferromagnetic order is reduced. As a consequence, the magnetic field induced hysteresis loops at I and III-quadrant washes out at 3K and bifurcation temperature between FC and ZFC shifted towards 24K anomaly in the derivatives. However, $T_N$ of $CuMn_{1.50}Ti_{0.50}O_4$ coincides exactly with the 24K anomaly followed by completely flattened FC and non-exponentially decrease of ZFC on cooling are the clear indication for the formation of spin-glass state.

In the above, the significant change in magnetization by Ti-substitution for Mn in $CuMn_2O_4$ is explained using Curie-weiss law and Arrott plot which are suggesting the materials as inhomogeneous system. But, the inconsistent change in magnetization of the derivatives at room temperature requires to explore further. So, in-order to perceive this, the room temperature dielectric spectroscopy measurement is discussed in the next section.

### 3.3 Dielectric Spectroscopy Measurements

#### *3.3.1 Complex dielectrics and AC conductivity*

The frequency dependent complex dielectric permittivity of $CuMn_{2-x}Ti_xO_4$ was measured at 300K and subsequently plotted the real part of complex dielectric permittivity ($\varepsilon'$) and its dielectric loss ($\tan\delta = \varepsilon''/\varepsilon'$) using semi-log scale in Fig. 7(a) and (b) respectively. Fig. 7(a) clearly shows the exponential decrease in $\varepsilon'$ with increase in frequency of $CuMn_2O_4$ implies the time-dependent relaxation. This is completely different compared to our previous result as expected due to the purity and different cationic distributions [25]. Further, the very weak frequency dependent of $\varepsilon'$ in $CuMn_{1.75}Ti_{0.25}O_4$ and $CuMn_{1.50}Ti_{0.50}O_4$ suggests a phenomenon of nearly constant loss which results from the relaxation involving with the motion of highly localized charges rather than dominating hopping process [26]. Generally, the exponential decreasing behaviour in dielectric permittivity with frequency is explained by Koop's theory for an inhomogeneous medium of two layers of the Maxwell-Wagner type in which, the grains and grain boundaries are the constituents [27, 28]. But, inset fig. 7(b) shows that the dielectric loss ($\tan\delta$) of $CuMn_2O_4$ decreases exponentially below 10kHz frequency above which it continues as independent of frequency up to the higher limit quite similar with the previous result [25]. Upon substituting Ti for Mn in $CuMn_2O_4$, fig 7(b) shows the faster decrease in $\tan\delta$ as compared to pure one at low frequency. However, the $\tan\delta$ value of $CuMn_{1.50}Ti_{0.50}O_4$ is significantly reduced compared to $CuMn_{1.75}Ti_{0.25}O_4$ below 10kHz may be due to the presence of $TiO_2$ which increases the grain boundary effect and thus the low conducting phase of grain boundaries enhances the energy of electron to exchange.

In order to understand the conduction mechanism on these materials, the AC conductivity is obtained from real part of dielectric permittivity ($\varepsilon'$) and dielectric loss (tanδ) with respect to frequency as given below,

$$\sigma_{AC} = 2\pi\varepsilon_0\varepsilon' f \tan\delta \qquad (2)$$

Here, $\varepsilon_0$ is the dielectric permittivity of free space and $f$ is the frequency of the applied field. Fig. 8 clearly shows a linear variation in AC conductivity with frequency. Generally, the linear behaviour of AC conductivity with frequency in disordered materials behaves as $\sigma_{AC} \sim \omega^s$ which implies the hopping conduction. Here, $s$ is the exponent of frequency and defined as $s = 1 - (6k_B T / W_m)$ where $W_m$ is the maximum barrier height [29, 30]. Analogous to our previous result, the conductivity of $CuMn_2O_4$ decreases with decrease in frequency down to 10kHz reveals that $s > 0$ implies $6k_B T < W_m$. But below 10kHz, conductivity becomes almost independent of frequency. A similar type of frequency independent conductivity with increased values are obtained for $CuMn_{2-x}Ti_xO_4$ which provides $s \approx 0$ implies $6k_B T \approx W_m$ within the measurement region associated with increase of grain boundary effect which causes short-range hopping conduction. However, the magnitude of conductivity of $CuMn_{1.50}Ti_{0.50}O_4$ is less as compared to $CuMn_{1.75}Ti_{0.25}O_4$ may be due to the effect of $TiO_2$ which behaves as the grain boundary. Further, the behaviour of $\sigma_{AC}$ within 0.1 – 1 kHz frequency range is expected to be associated with the itinerant electron magnetism causes weak ferromagnetic. The cause of increase in conductivity in $CuMn_{2-x}Ti_xO_4$ can be understood from impedance and electric modulus spectroscopy in the next sub-section.

*3.3.2 Impedance and Electric modulus*

The room temperature frequency dependence of $Z'$ and $Z''$ for $CuMn_{2-x}Ti_xO_4$ are depicted in fig. 9(a) and (b) respectively using semi-log scale. In contrast to our previous impedance results of $CuMn_2O_4$, here, the obtained $Z'$ is quite similar to that but $Z''$ shows a broad peak at higher frequency side instead of the inflection point around 10kHz [25]. In $CuMn_{1.75}Ti_{0.25}O_4$, the magnitude of $Z'$ reduces and peak position of $Z''$ shifts towards higher limit of frequency associated with reduced magnitude as compared to $CuMn_2O_4$. But, $Z'$ and $Z''$ of $CuMn_{1.50}Ti_{0.50}O_4$ associated with impurity phase $TiO_2$ has less variation as compared to $CuMn_{1.75}Ti_{0.25}O_4$ and lies close to $CuMn_2O_4$. Here, the reduced magnitude of $Z'$ is expected to be due to the decrease of immobile charges and the shifting of peak position in $Z''$ implies that the relaxation depends strongly on chemical compositions. Further, the decreasing trend of $Z'$ at higher frequency limit is drastically reduced may be due to the increasing barrier effect in the derivatives. In order to distinguish the contribution of grain, grain boundary and electrode effects at room temperature, the complex impedance for $CuMn_{2-x}Ti_xO_4$ are analyzed by Nyquist plot ($Z''$ vs. $Z'$) as can be seen from fig. 10. Within the range of measurement frequency, the Nyquist plot reveals the distinct effect of grain boundary associated with a single semi-circle at intermediate frequency range because higher frequency comprises with grain and lower frequency comprises with electrode effects. Further, all the semicircles are asymmetric in Nyquist plot reveals the non-Debye type relaxation behaviour. Generally, the impedance in complex dielectric spectroscopy is due to only resistive components. Therefore, the contribution of capacitive reactance in $CuMn_{2-x}Ti_xO_4$ is analysed by electric modulus data as follows.

Fig. 11(a) and (b) shows the frequency dependence of $M'$ and $M''$ respectively in semi-log scale for $CuMn_{2-x}Ti_xO_4$ at room temperature. In $CuMn_2O_4$, the $M'$ and $M''$ behaviour are quite similar with the reported result except the inflection point which is obtained as peak at higher frequency side. However, both the figures show similar type of frequency dependence for all the materials with very weak dependent of frequency below 10 kHz above which they rapidly increase with significant difference between them. This implies that the hopping conduction process is dominated by short-range mobility of charge carriers for all the materials. In addition, we have also plotted $M''$ vs. $M'$ for all the materials at room temperature as shown in fig. 12 in which $CuMn_2O_4$ shows two semi-circles one at lower frequency range associated with grain boundary effect and other at higher frequency range associated with grain effect quite similar to previous result [25]. But, only one semi-circle is obtained in the derivatives with increase in size within the measurement frequency range due to the increasing effect of capacitive grain boundary.

## 4 Conclusions

In the present work, we have reported the magnetization and dielectric spectroscopy measurements on tetragonal symmetry of polycrystalline $CuMn_{2-x}Ti_xO_4$ which is accompanied with small amount of $Mn_3O_4$ as impurity phase obtained by the Rietveld analysis of the room temperature neutron diffraction. The formation of phase is also confirmed by FTIR spectroscopy. Further, the presence of $TiO_2$ as un-reacted phase in $CuMn_{1.50}Ti_{0.50}O_4$ has significant influence on magnetic and dielectric behaviour at room temperature. In $CuMn_2O_4$, the effective magnetic moment '$\mu_{eff}$' is obtained less compared to our previous result may be due to the purity and different cationic distribution of magnetic ions which are also responsible for the decrease of $\mu_{eff}$ in $CuMn_{2-x}Ti_xO_4$ as expected. But, however, the large $\mu_{eff}$ of $CuMn_{1.50}Ti_{0.50}O_4$ as compared to $CuMn_{1.75}Ti_{0.25}O_4$ below 175K is expected to be due to the presence of un-reacted $TiO_2$ phase. Upon Ti-substitution, the magnetic field induced hysteresis loop gradually washes out at low temperature and the anti-ferromagnetic order renormalizes the ferromagnetic order in such a way that $CuMn_{1.50}Ti_{0.50}O_4$ reveals the spin-glass state. Because, the competition between anti-ferromagnetic and ferromagnetic ordering in a spin-glass state results completely flattened FC and non-exponentially decrease in ZFC below the bifurcation temperature. But, the room temperature weak ferromagnetic significantly reduces in $CuMn_{1.50}Ti_{0.50}O_4$ as compared to $CuMn_{1.75}Ti_{0.25}O_4$ which is close to $CuMn_2O_4$ in the Arrott's plot. Further, the room temperature complex dielectric, impedance and electric modulus reveals the electronic conduction by hopping associated with short-range mobility of charge carriers in $CuMn_2O_4$. The same conduction process with increasing grain boundary effect was obtained upon substituting Ti for Mn in $CuMn_2O_4$. But, the presence of $TiO_2$ un-reacted phase in $CuMn_{1.50}Ti_{0.50}O_4$ reverts back close to $CuMn_2O_4$ behavior in magnetization, AC conductivity, impedance and electric modulus at room temperature.

**Acknowledgement**

One of the authors (I. Naik) acknowledges UGC-DAE CSR, Mumbai Centre for the experimental facility under the collaborative research project scheme (CRS-M-229). P. Patra would like to acknowledge for the fellowship under this project. All the authors would like to express their gratitude to Dr. S. Mukherjee, UGC-DAECSR Mumbai Centre and Dr. B. S. Mohanta, Lab. Technician, Dept. of Chemistry, NOU for the Dielectric and FTIR measurements respectively.

## FIGURE CAPTIONS

**Figure 1:** (Colour online) Left side figure shows the Rietveld refinement of $CuMn_{2-x}Ti_xO_4$ (x = 0. 0.25, 0.5) neutron diffraction patterns measured at 300K. Right side figure shows the reduction of $Mn_3O_4$ concentration in the doublet peak.

**Figure 2:** (Colour online) Room temperature FTIR spectroscopy of $CuMn_{2-x}Ti_xO_4$.

**Figure 3:** (Colour online) Temperature dependent magnetization of $CuMn_{2-x}Ti_xO_4$. Inset figure shows temperature dependent inverse susceptibility of $CuMn_{2-x}Ti_xO_4$ indicating the ferrimagnetic state.

**Figure 4:** (Colour online) Normalized susceptibility of $CuMn_{2-x}Ti_xO_4$ in the temperature range of 2–100K. Inset figure shows the Curie-Weiss fitting of $CuMn_{2-x}Ti_xO_4$.

**Figure 5:** (Colour online) Hysteresis loop in I-quadrant of magnetic field dependent magnetization on $CuMn_{2-x}Ti_xO_4$ at 3K and 300K. Inset figure shows the M(H) behaviour of $CuMn_{2-x}Ti_xO_4$ at 300K.

**Figure 6:** (Colour online) Room temperature Arrott plot of $CuMn_{2-x}Ti_xO_4$ after subtraction of van-vleck magnetization. Inset figure shows Arrott plot of $CuMn_{2-x}Ti_xO_4$ using raw data.

**Figure 7:** (Colour online) Frequency dependent (a) real part of complex permittivity and (b) dielectric loss of $CuMn_{2-x}Ti_xO_4$ at room temperature. Inset fig. 7(b) shows the dielectric loss of $CuMn_2O_4$.

**Figure 8:** (Colour online) Frequency dependent AC conductivity of $CuMn_{2-x}Ti_xO_4$ at room temperature.

**Figure 9:** (Colour online) Frequency dependent (a) real part of complex impedance and (b) imaginary part of complex impedance of $CuMn_{2-x}Ti_xO_4$ at room temperature.

**Figure 10:** (Colour online) Room temperature Nyquest plot of $CuMn_{2-x}Ti_xO_4$.

**Figure 11:** (Colour online) Frequency dependent (a) real part of complex electric modulus and (b) imaginary part of complex electric modulus of $CuMn_{2-x}Ti_xO_4$ at room temperature.

**Figure 12:** (Colour online) Room temperature electric modulus plot of $CuMn_{2-x}Ti_xO_4$.

Table – I

Magnetic parameters for $CuMn_{2-x}Ti_xO_4$ (x = 0.0, 0.25 and 0.50)

| Compounds | $CuMn_2O_4$ | $CuMn_{1.75}Ti_{0.25}O_4$ | $CuMn_{1.50}Ti_{0.50}O_4$ |
|---|---|---|---|
| Lattice Parameter (Å) | a = 5.871(3) | a = 5.894(6) | a = 5.931(2) |
| | c = 8.313(5) | c = 8.284(4) | c = 8.279(3) |
| Cell volume (Å$^3$) | 286.522(6) | 287.438(9) | 291.206(5) |
| $\chi_0(emu/Oe.mole)$ | 0.0045 | 0.018 | 0.0067 |
| $T_N(K)$ | 56 | 51 | 25 |
| $\theta_{CW}(K)$ | 75 | 70 | 70 |
| $C(emuK/mole)$ | 1.841 | 1.615 | 1.757 |
| $\mu_{eff}(\mu_B)/f.u.$ | 3.79 | 3.56 | 3.71 |

Table – II

Summary of Disorder in $CuMn_{2-x}Ti_xO_4$ (x = 0.0, 0.25 and 0.50) as refined from the neutron diffraction data at 300K, with space group *I*41/*amd* where Cu is located at 4b (0 ¼ 3/8), Mn is located at 8c (0 0 0) and O is located at 16h (0 0.5226 0.2308) site.

| Materials | $CuMn_2O_4$ | $CuMn_{1.75}Ti_{0.25}O_4$ | $CuMn_{1.50}Ti_{0.50}O_4$ |
|---|---|---|---|
| Cu/Mn/Ti | 0.92/0.08 | 0.84/0.13/0.03 | 0.90/0.07/0.03 |
| Mn/Cu/Ti | 0.94/0.06 | 0.78/0.11/0.11 | 0.72/0.10/0.18 |

**Figure 1**

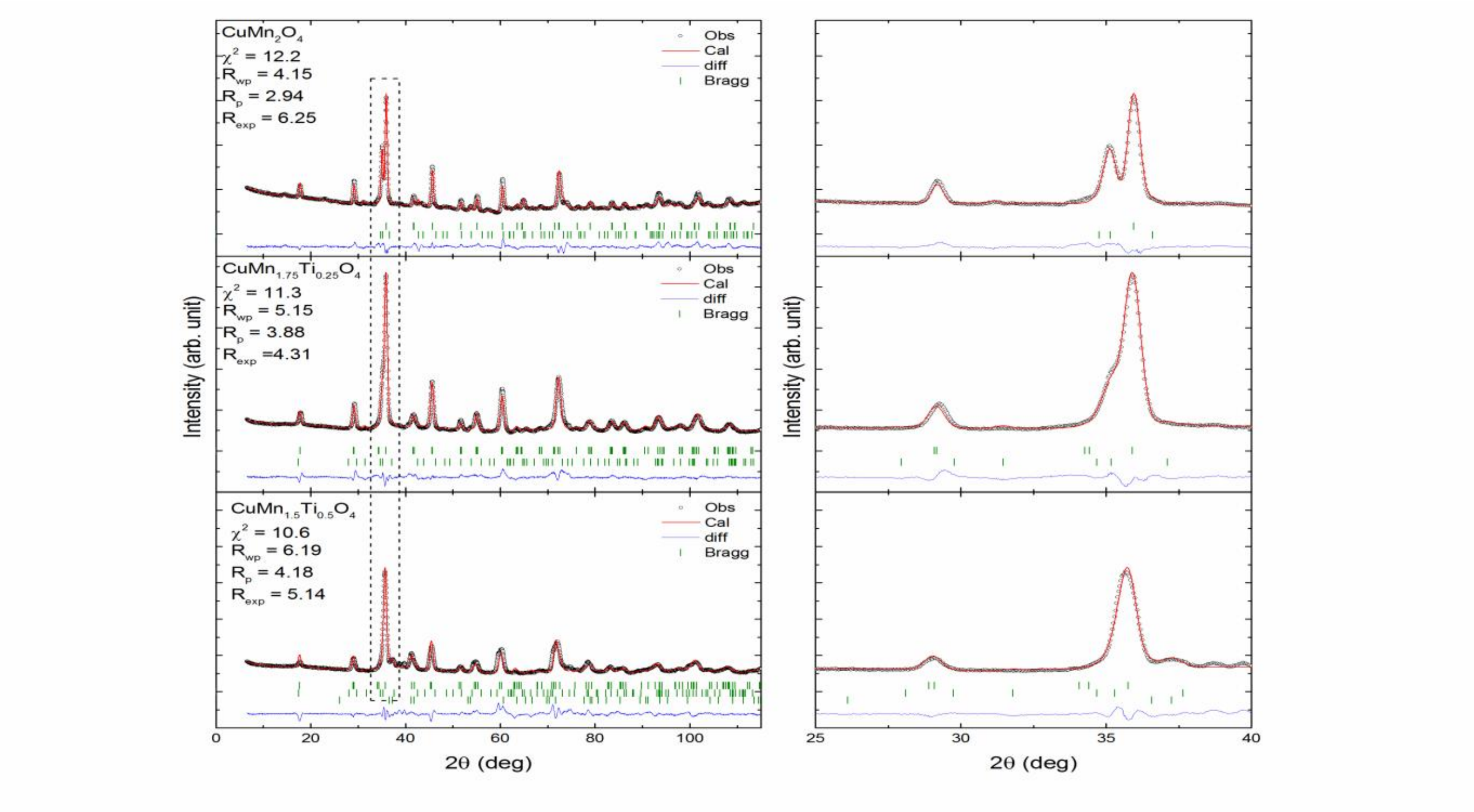

**Figure 2**

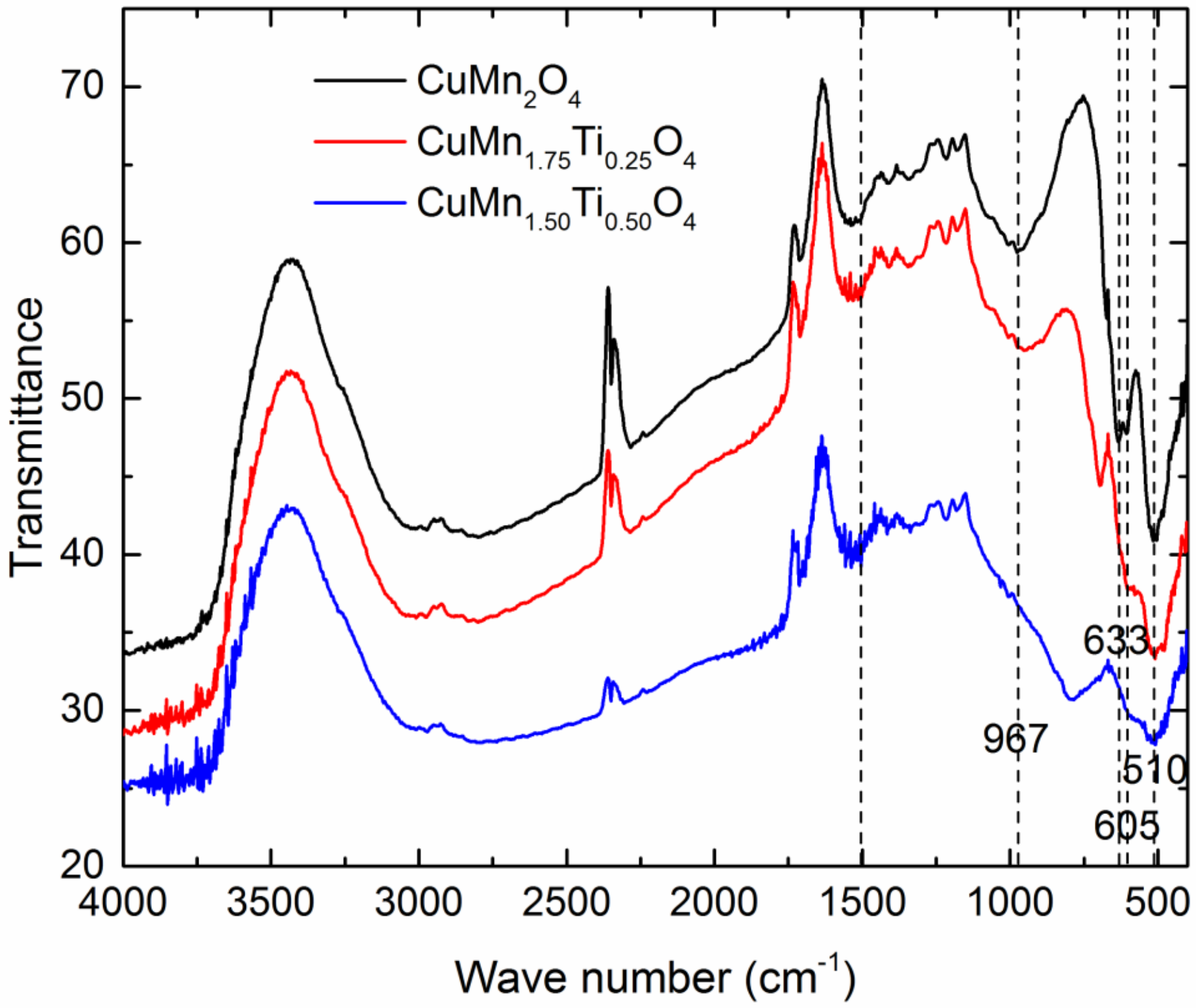

**Figure 3**

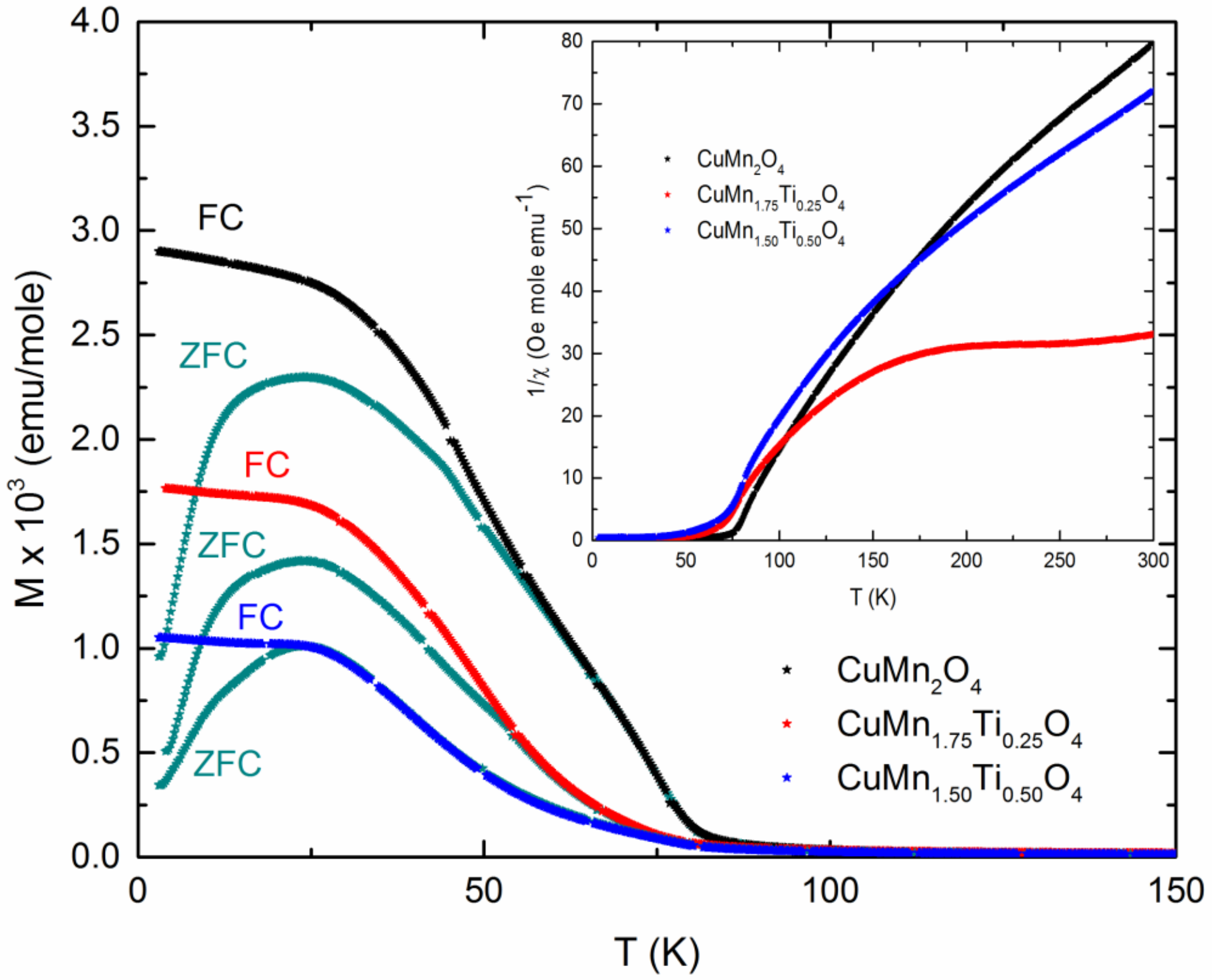

**Figure 4**

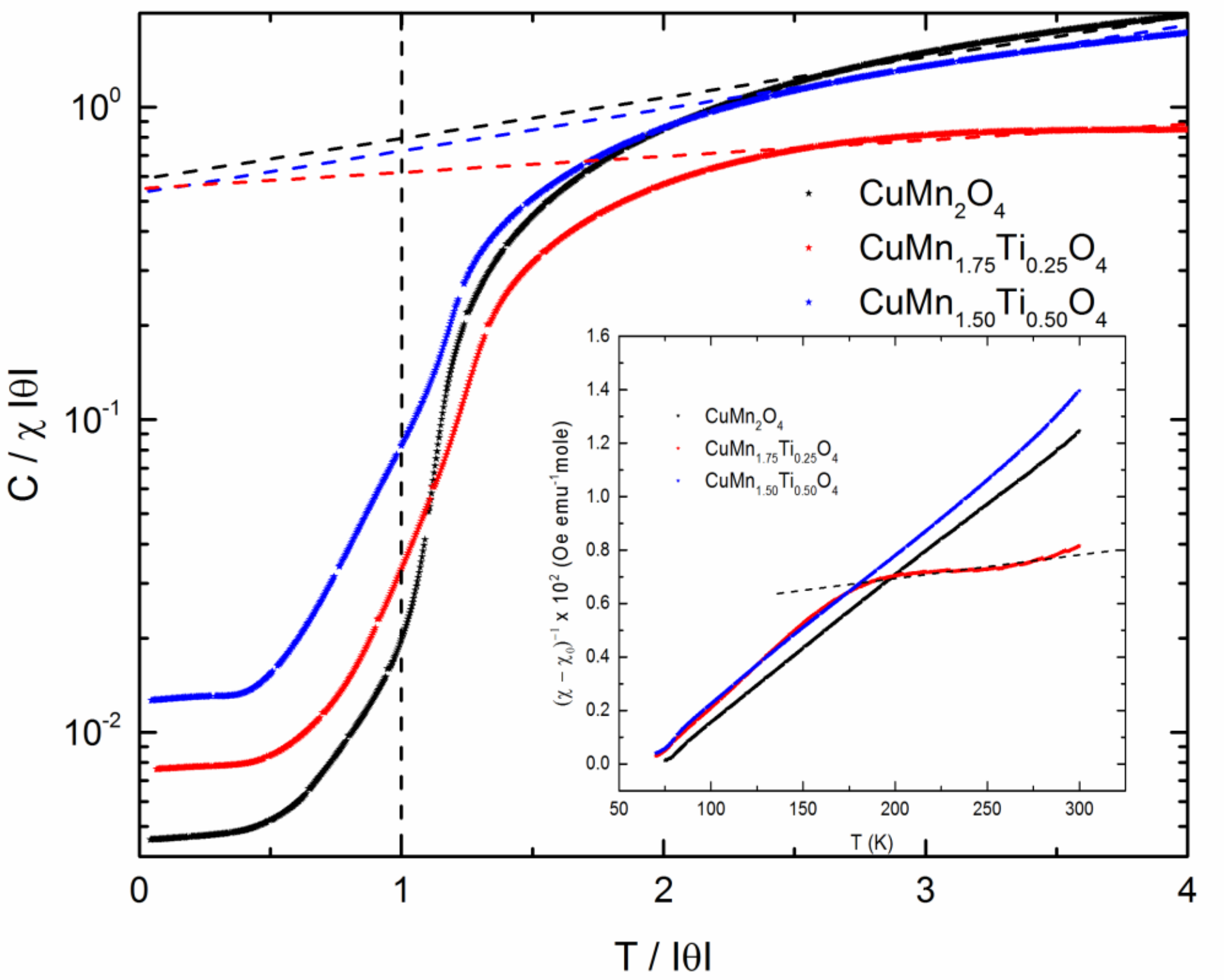

**Figure 5**

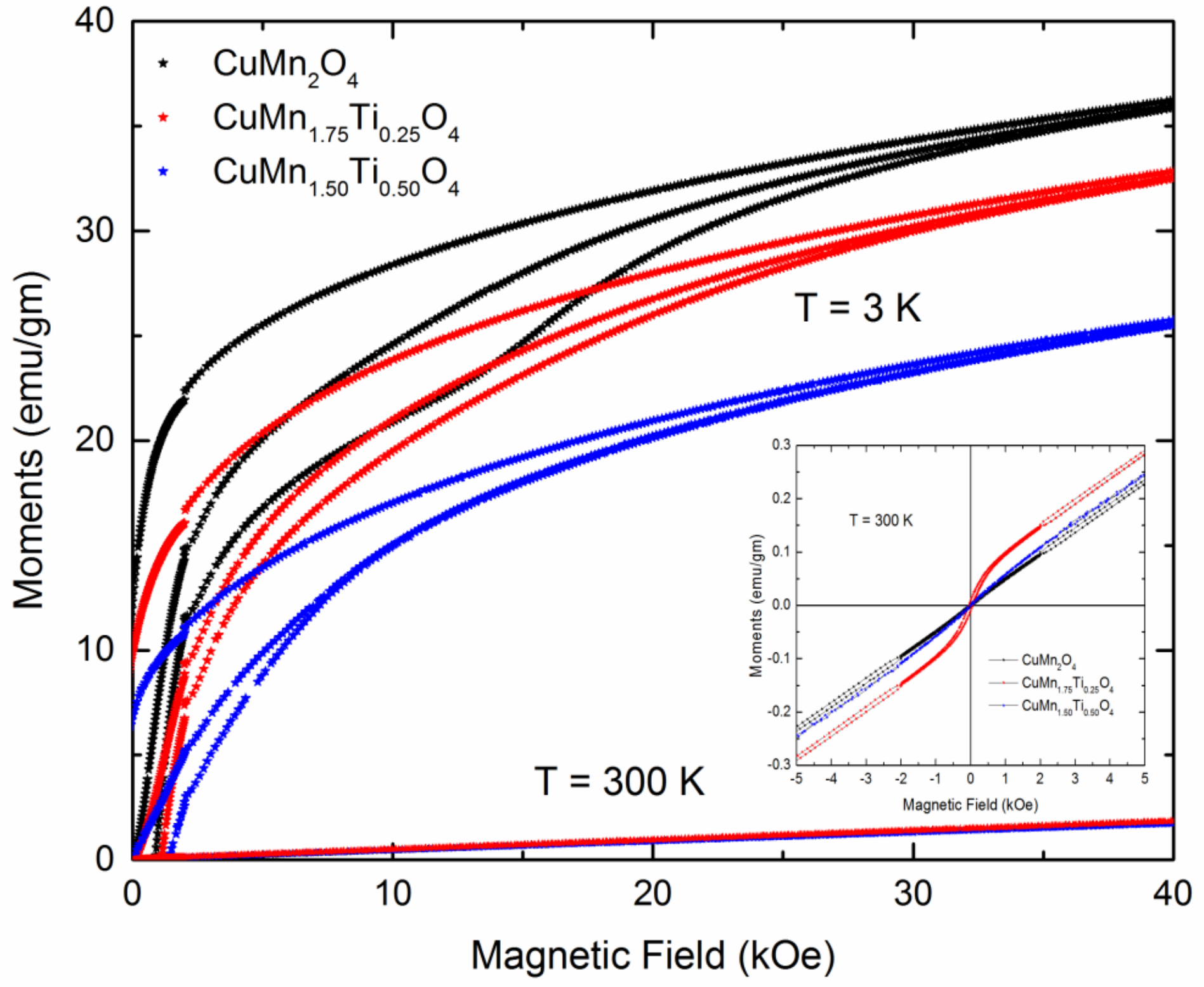

**Figure 6**

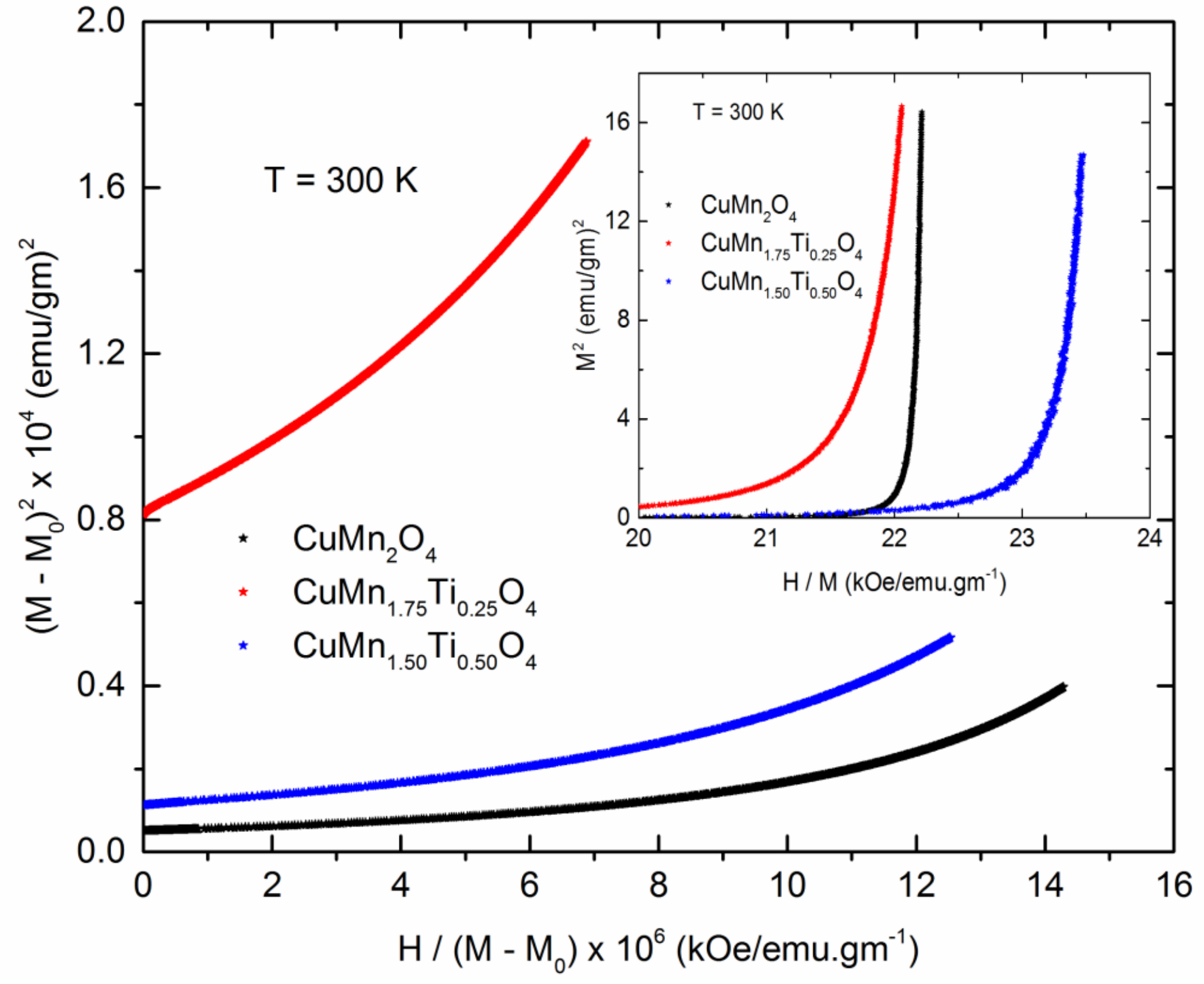

**Figure 7**

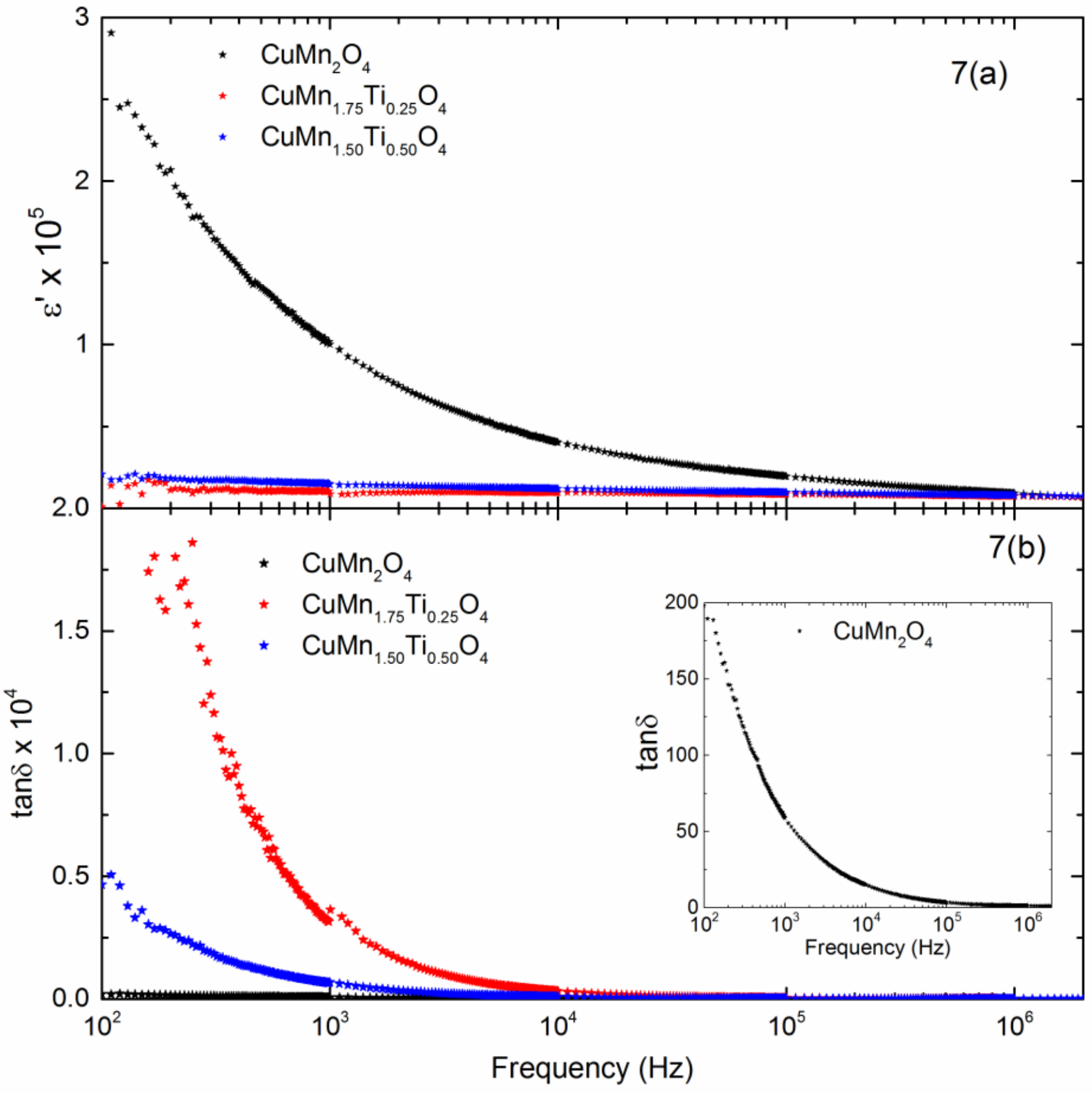

**Figure 8**

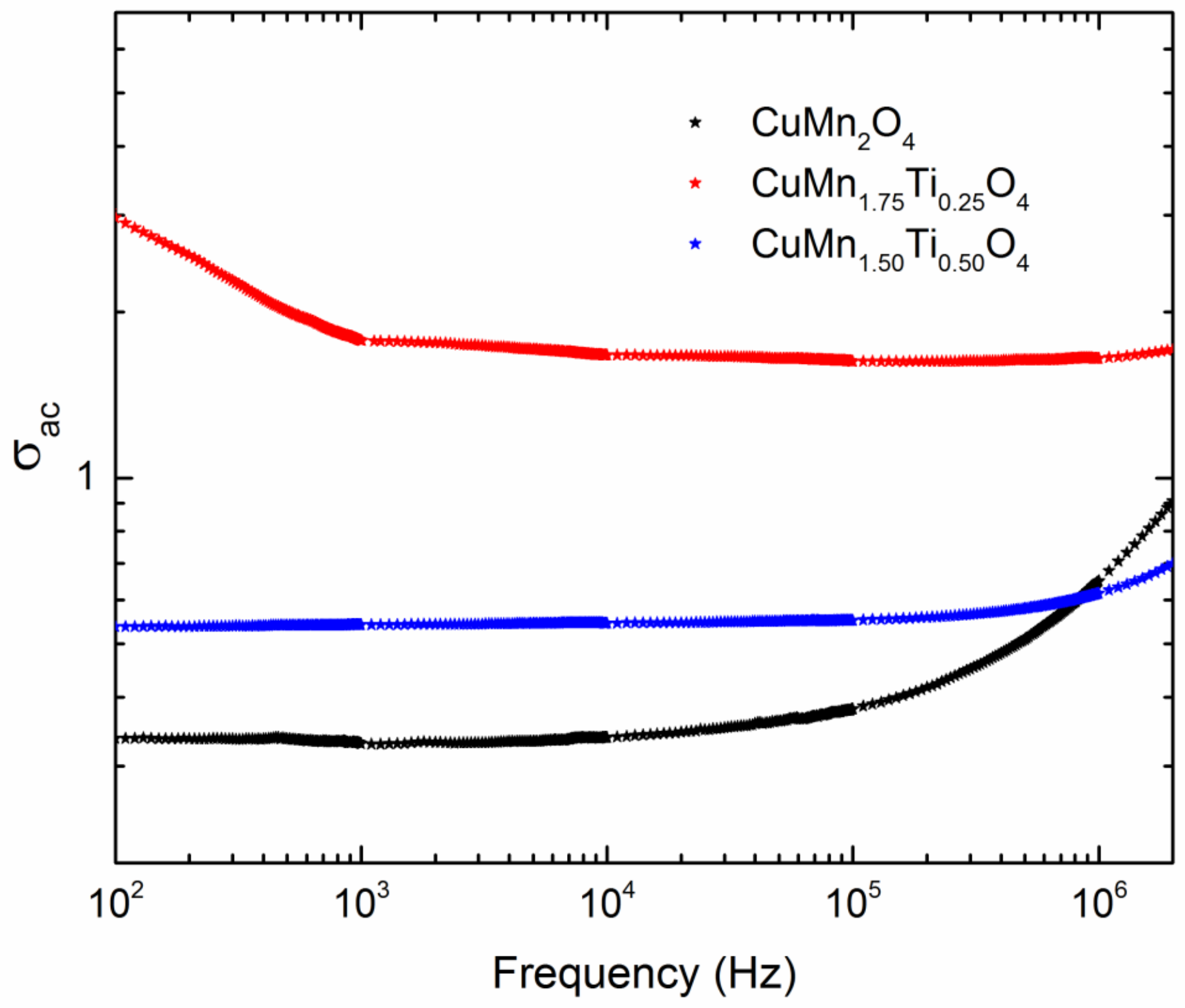

**Figure 9**

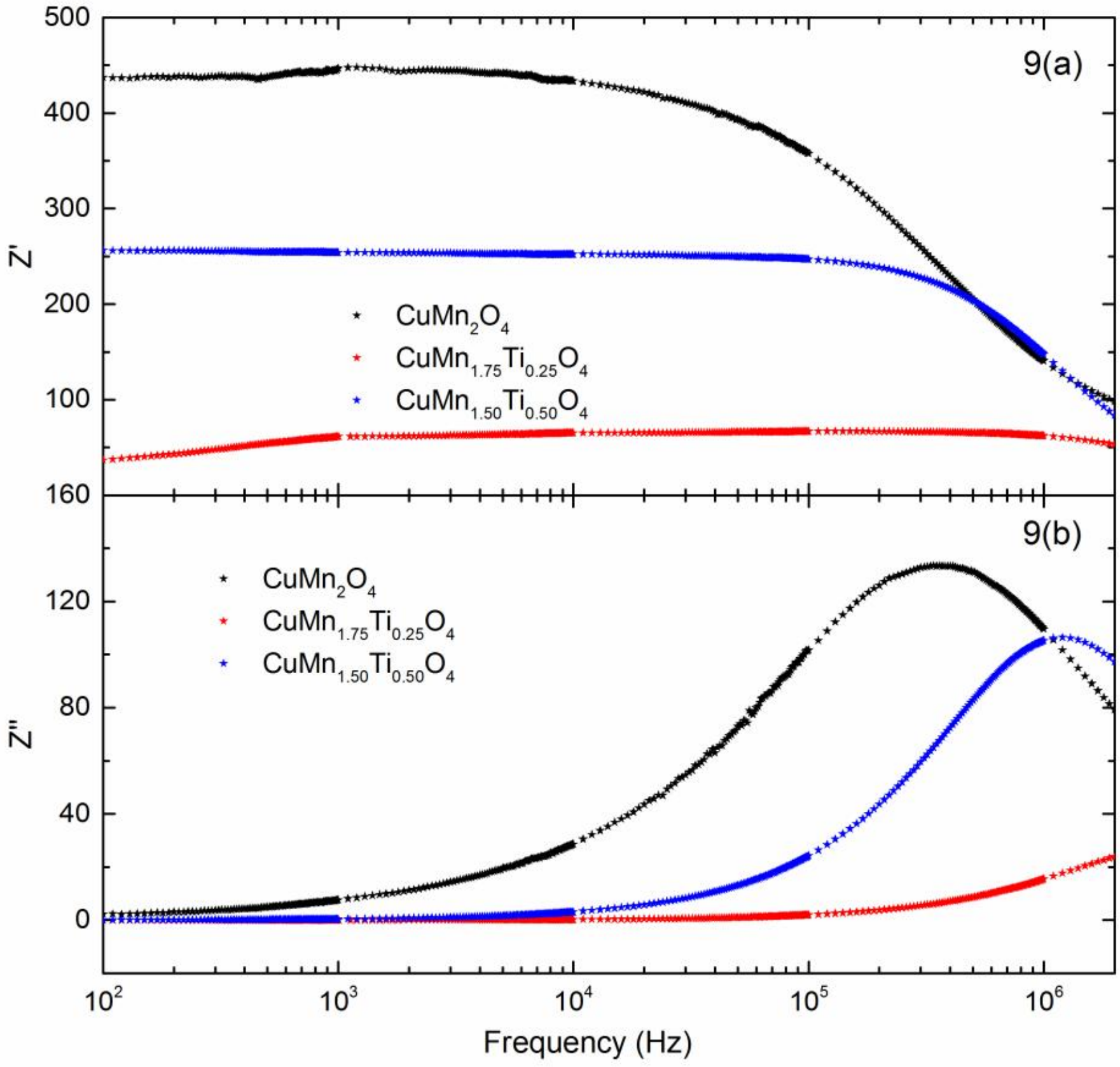

**Figure 10**

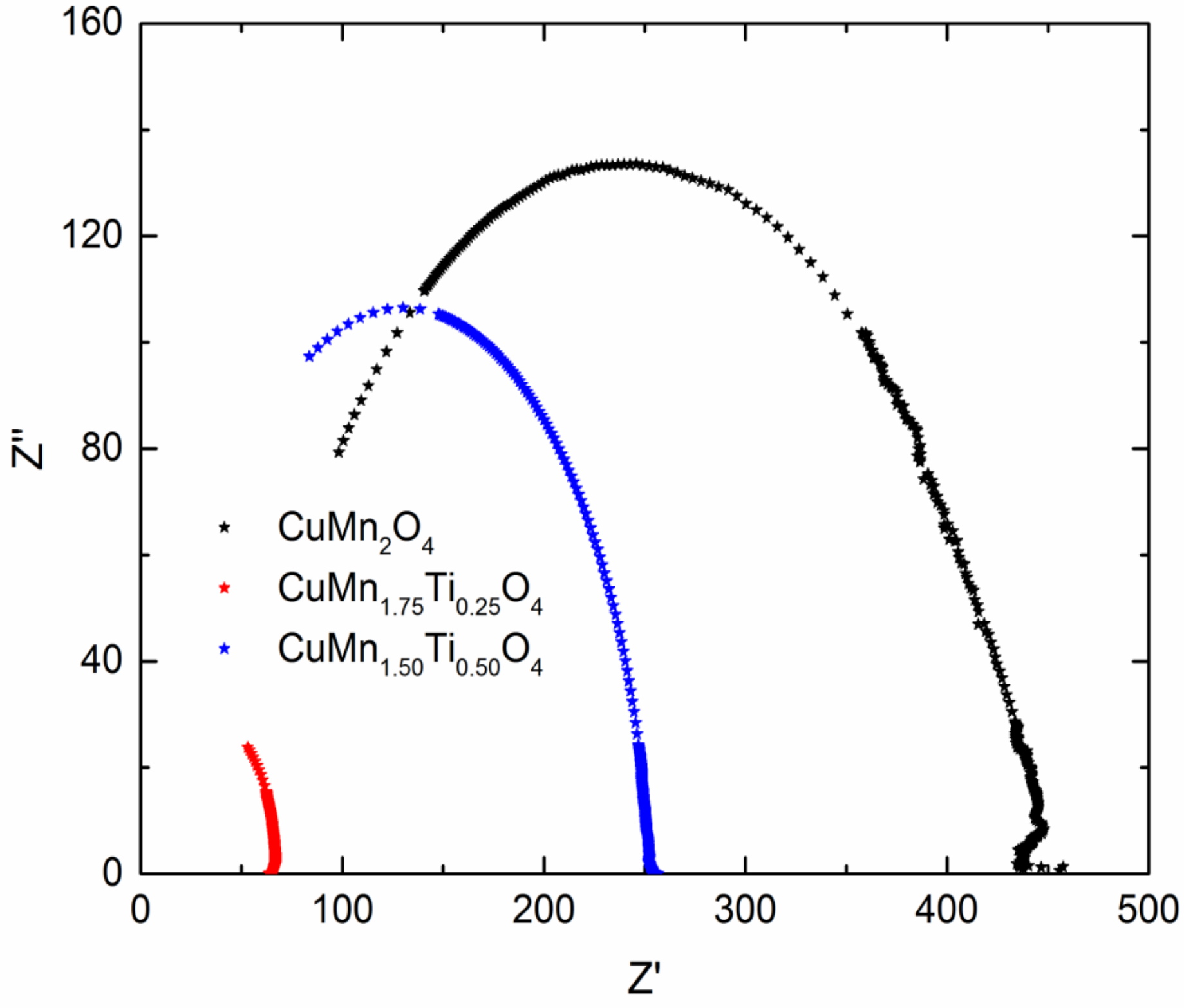

**Figure 11**

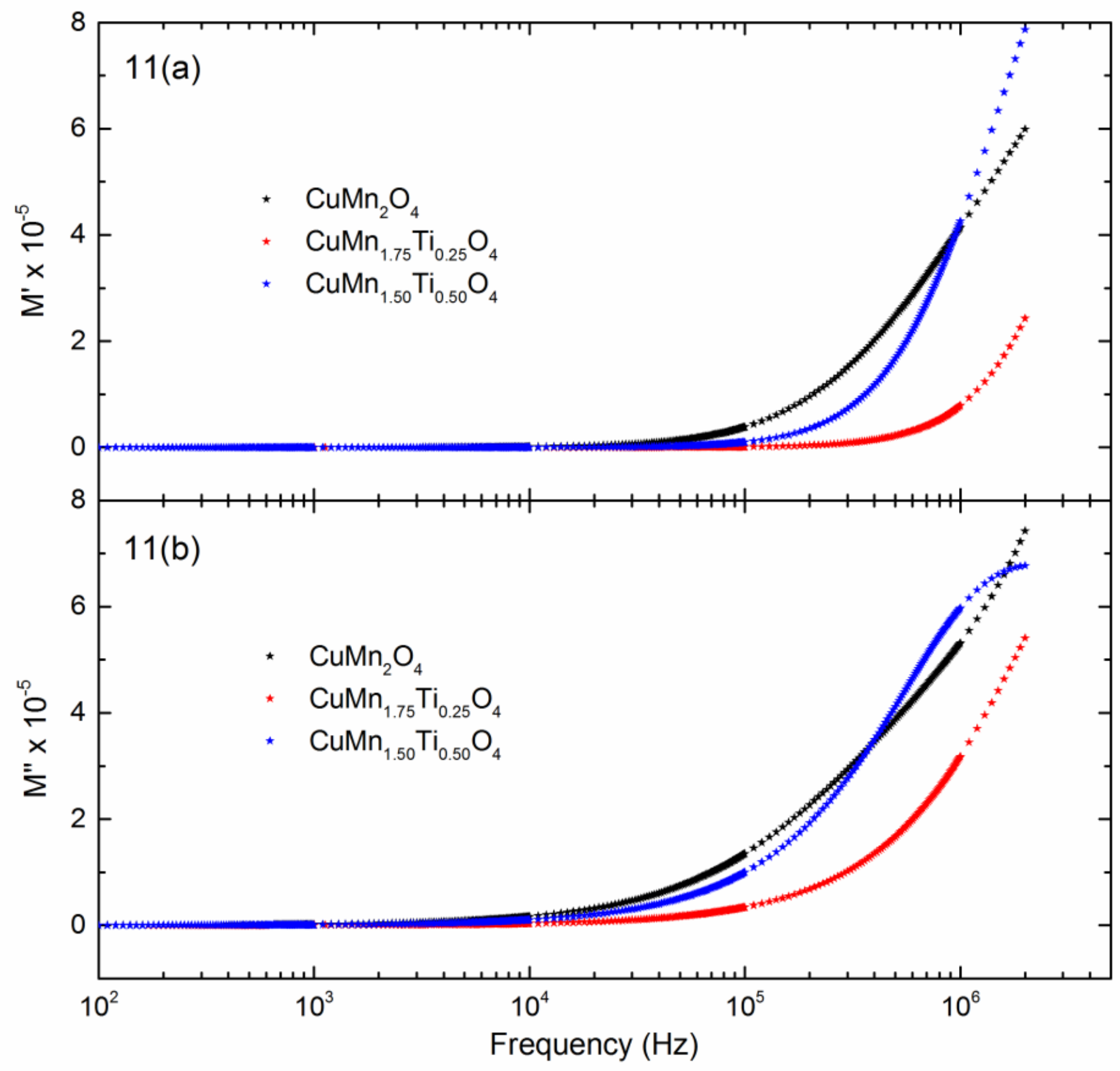

**Figure 12**

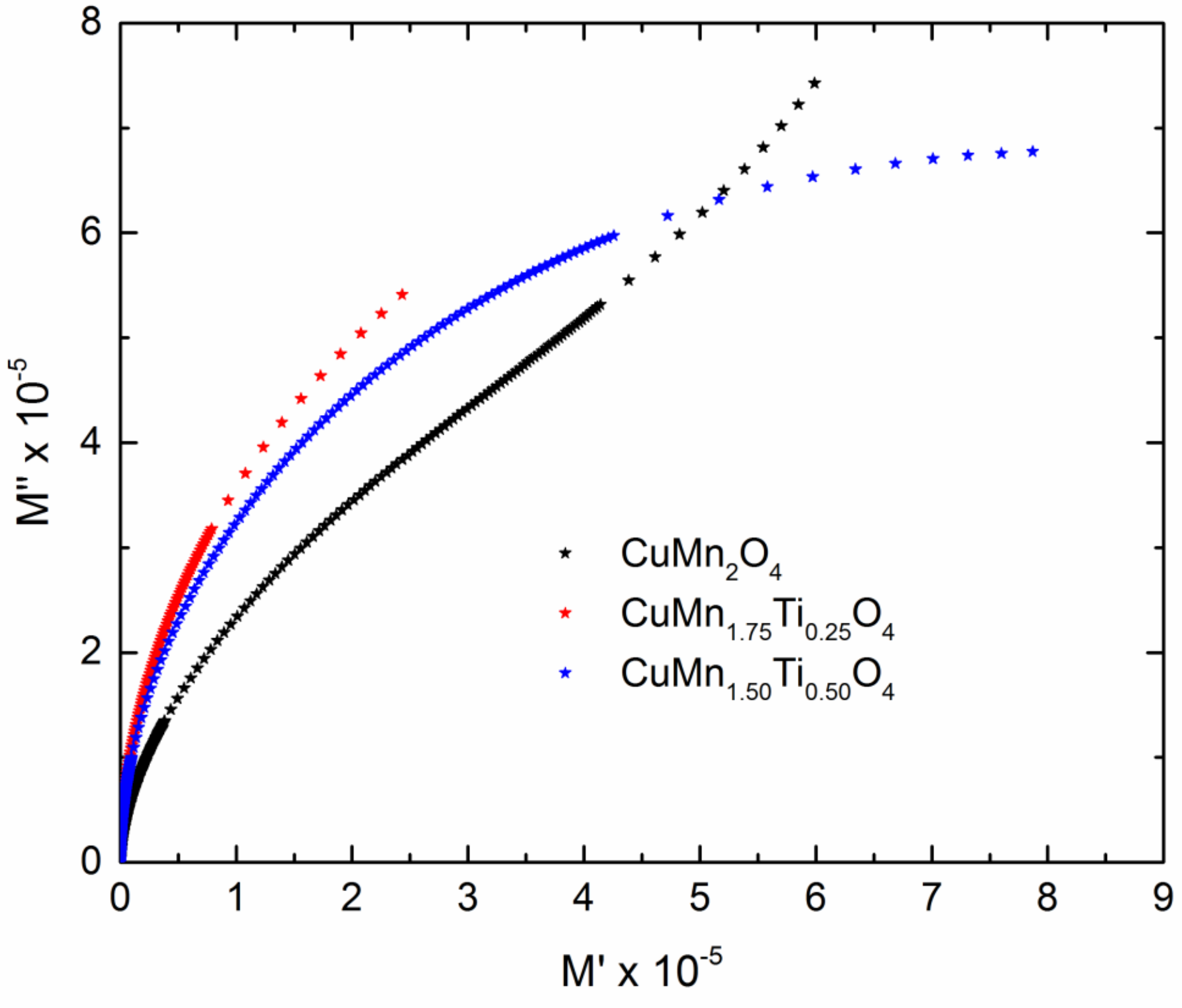